\documentclass[aps,prl,reprint,superscriptaddress,showpacs,floatfix,longbibliography]{revtex4-1}

\usepackage{graphicx}
\usepackage{amssymb,amsmath}
\usepackage{dcolumn}
\usepackage{bm}
\usepackage[mathlines]{lineno}
\usepackage{soul,color}
\begin{document}

\title{Strain-controlled fundamental gap and structure of
       bulk black phosphorus}

\author{Jie~Guan}
\affiliation{Physics and Astronomy Department,
             Michigan State University,
             East Lansing, Michigan 48824, USA}

\author{Wenshen~Song}
\affiliation{Department of Physics,
              Washington University in St. Louis,
              St. Louis, Missouri 63130, USA}

\author{Li~Yang}
\affiliation{Department of Physics,
              Washington University in St. Louis,
              St. Louis, Missouri 63130, USA}

\author{David Tom\'{a}nek}
\email
            {tomanek@pa.msu.edu}%
\affiliation{Physics and Astronomy Department,
             Michigan State University,
             East Lansing, Michigan 48824, USA}

\date{\today} 

\begin{abstract}
We study theoretically the structural and electronic response of
layered bulk black phosphorus to in-layer strain. {\em Ab initio}
density functional theory (DFT) calculations reveal that the
strain energy and interlayer spacing display a strong anisotropy
with respect to the uniaxial strain direction. To correctly
describe the dependence of the fundamental band gap on strain, we
used the computationally more involved GW quasiparticle approach
that is free of parameters and superior to DFT studies, which are
known to underestimate gap energies. We find that the band gap
depends sensitively on the in-layer strain and even vanishes at
compressive strain values exceeding ${\approx}2$\%, thus
suggesting a possible application of black P in strain-controlled
infrared devices.
\end{abstract}



\maketitle


Layered bulk black phosphorus (BP), discovered only a century
ago~\cite{Bridgman14}, is a direct-gap semiconductor with an
observed fundamental band gap%
~\cite{{bulkbp},{Warschauer63},{Narita83},{Maruyama81}} of
$0.31-0.36$~eV. Its electronic response distinguishes BP as
favorable from other well-studied layered systems including
semimetallic graphite and transition metal dichalcogenides (TMDs)
such as MoS$_2$, which are indirect-gap semiconductors. Under
compression, bulk BP has displayed an interesting change in its
electronic and topological
properties~{\cite{{SWS:Pressure:AKAHAMA1986397},
{SWS:Dirac:PhysRevB.91.195319},
{SWS:Pressure:PhysRevLett.115.186403}}. Similar to bulk BP, a much
wider direct fundamental band gap is present in phosphorene
monolayers and few-layer systems, suggesting promising
applications in 2D semiconductor
electronics~\cite{{DT229},{DT230},{Li2014}}. Whereas it is now
well established that the gap displays a strong and anisotropic
response to in-layer strain in phosphorene monolayers and
few-layer systems~\cite{{DT229},{DT230},%
{SWS:Strain:PhysRevLett.112.176801},{Li2014},{Yang2014},%
{SWS:Strain:APL},{SWS:GW:PhysRevB.90.205421},%
{SWS:Strain:Ju2015109}}, no dependable data are available for the
corresponding response in the bulk system.


To fill in this missing information, we study theoretically the
structural and electronic response of layered bulk black
phosphorus to in-layer strain. Our {\em ab initio} density
functional theory (DFT) calculations reveal that the strain energy
and interlayer spacing display a strong anisotropy with respect to
the uniaxial strain direction. To correctly describe the
dependence of the fundamental band gap on strain, we used the
computationally more involved GW quasiparticle approach that is
free of parameters and superior to DFT studies, which are known to
underestimate gap energies. We find that the band gap depends
sensitively on the in-layer strain and even vanishes at
compressive strain values exceeding ${\approx}2$\%, thus
suggesting a possible application of black P in strain-controlled
infrared devices.

\begin{figure}[b]
\includegraphics[width=1.0\columnwidth]{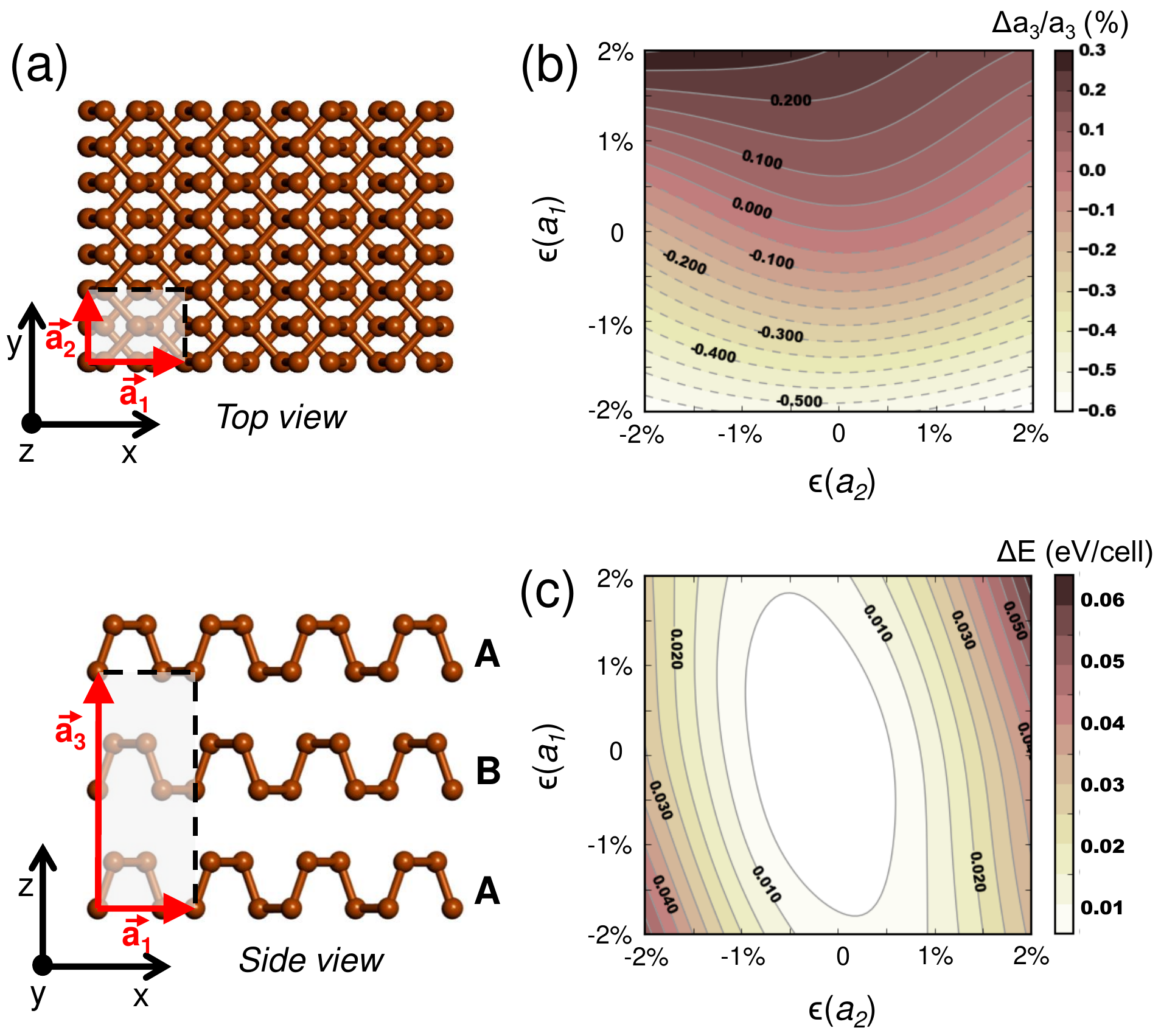}
\caption{(Color online) (a) Ball-and-stick model of the structure
of bulk black phosphorus in top and side views. (b) Fractional
change of the interlayer distance $a_3$ as a function of the
in-layer strain $\epsilon$ along the $a_1$ and $a_2$ directions.
(c) Dependence of the strain energy ${\Delta}E$ per unit cell on
the in-layer strain along the $a_1$ and $a_2$ directions.
\label{fig1} }
\end{figure}

\begin{figure*}[t]
\includegraphics[width=1.8\columnwidth]{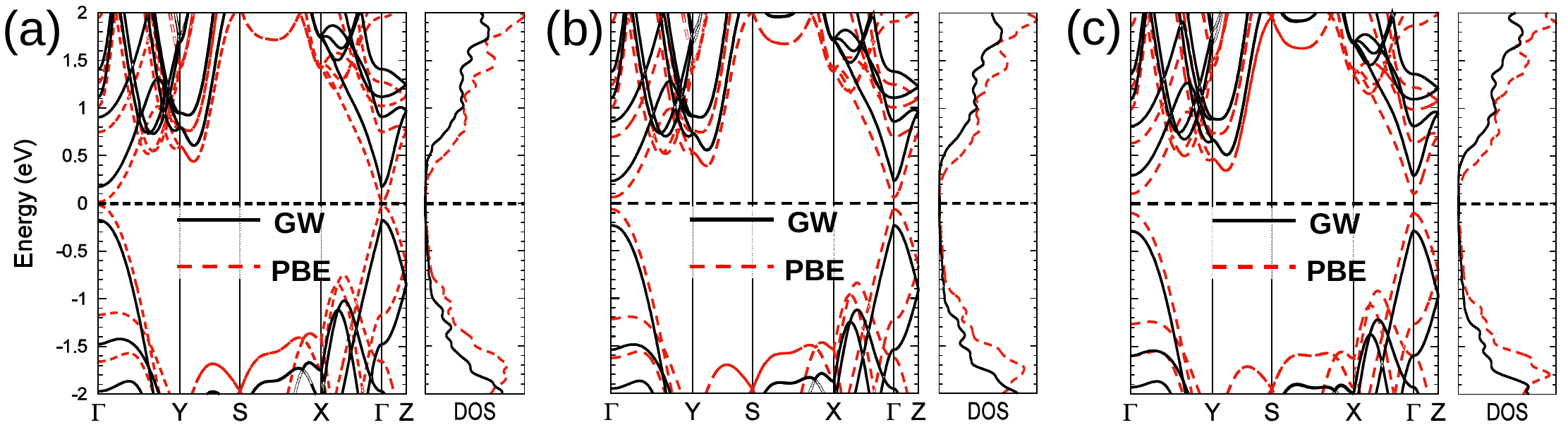}
\caption{(Color online) Electronic band structure (left panels)
and density of states (right panels) of bulk black phosphorus (a)
without strain, (b) when stretched by 1\% along the $a_1$
direction, and (c) when stretched by 2\% along the $a_1$
direction. GW results are shown by the solid black lines. DFT-PBE
results, which underestimate the band gap, are shown by the dashed
red lines.%
\label{fig2} }
\end{figure*}


We utilize {\em ab initio} density functional theory (DFT) as
implemented in the {\textsc SIESTA}~\cite{SIESTA} code to optimize
the structure and to determine the structural response to in-plane
strain. We used the Perdew-Burke-Ernzerhof (PBE)~\cite{PBE}
exchange-correlation functional, norm-conserving Troullier-Martins
pseudopotentials~\cite{Troullier91}, and a double-$\zeta$ basis
including polarization orbitals. The reciprocal space was sampled
by a fine grid~\cite{Monkhorst-Pack76} of $8{\times}8{\times}4$
$k$-points in the first Brillouin zone of the primitive unit cell
containing 8 atoms. We used a mesh cutoff energy of $180$~Ry to
determine the self-consistent charge density, which provided us
with a precision in total energy of ${\leq}2$~meV/atom. All
geometries have been optimized using the conjugate gradient
method~\cite{CGmethod}, until none of the residual
Hellmann-Feynman forces exceeded $10^{-2}$~eV/{\AA}.

DFT calculations are not designed to reproduce the electronic band
structure correctly. Even though DFT band structure usually
resembles observed results, the fundamental band gap is usually
underestimated. The proper way to calculate the band structure
without adjustable parameters involves solving the self-energy
equation. We perform such calculations using the GW
approximation~\cite{GW86} as implemented in the BerkeleyGW
package~\cite{BerkeleyGW}, where the dynamical electronic
screening is captured by the general plasmon pole
model.~\cite{GW86} We prefer this state-of-the-art approach to
computationally less involved hybrid DFT functionals such as
HSE06~\cite{HSE06}, which mix Hartree-Fock and DFT-PBE
exchange-correlation energies using an adjustable parameter. We
use single-shot $G_0W_0$ calculations with a
$14{\times}10{\times}4$ $k$-point grid, which provide converged
results for the self-energies and quasiparticle energy gaps in
strained bulk black phosphorus.


The optimum structure of bulk black phosphorus, as obtained by
DFT-PBE calculations, is shown in Fig.~\ref{fig1}(a). As seen in
the bottom panel, individual layers in the layered structure are
not flat due to the non-planar $sp^3$ hybridization of the P
atoms. The AB stacking is caused by displacing every other layer
along the $\vec{a}_2$-direction, yielding an orthorhombic lattice
spanned by the orthogonal lattice vectors $\vec{a}_1$, $\vec{a}_2$
and $\vec{a}_3$, with $\vec{a}_3$ extending over two inter-layer
distances. The covalent in-plane bonding is adequately described
by DFT-PBE, as suggested by the agreement between the calculated
lattice constants, $a_1({\rm PBE})=4.53$~{\AA} and $a_2({\rm
PBE})=3.36$~{\AA}, and the experimental
values~\cite{redp-blackp-phase1} $a_1({\rm expt})=4.38$~{\AA} and
$a_2({\rm expt})=3.31$~{\AA}. As indicated by recent Quantum Monte
Carlo studies~\cite{DT250}, the nature of the weak inter-layer
interaction in bulk black phosphorus differs in a non-trivial
manner from a van der Waals interaction. In view of this fact, the
calculated value of the out-of-plane lattice constant $a_3({\rm
PBE})=11.15$~{\AA} agrees rather well with the observed
value~\cite{redp-blackp-phase1} $a_3({\rm expt})=10.50$~{\AA}.

As a result of the weak inter-layer interaction contrasting the
strong in-layer covalent bonding, we do not expect the interlayer
distance to change much when the lattice is subjected to in-layer
strain. We considered both tensile and compressive in-layer strain
$\epsilon$ up to 2\%. Results for the fractional change
${\Delta}a_3/a_3$ for different strain combinations
${\epsilon}(a_1),{\epsilon}(a_2)$ are presented in
Fig.~\ref{fig1}(b). The continuous contour plot is based on a
cubic spline interpolation of a $5{\times}5$ grid of data points
for different strain value combinations. These results, as well as
our findings, suggest that the interlayer spacing changes much
less than 1\% for the strain range considered here. Such small
changes in the interlayer distance are unlikely to be affected by
the selection of the total energy functional, which plays only a
minor role in the interlayer separation~\cite{DT250} and are
consistent with very weakly coupled layers.

Our results in Fig.~\ref{fig1}(b) indicate a trend that the
inter-layer distance increases by stretching and decreases by
compressing the crystal along the soft, accordion-like $a_1$
direction. In contrast, both stretching and compression along the
stiffer $a_2$ direction cause a reduction of the inter-layer
spacing. Even though these effects are small, they clearly reflect
the anisotropy of the system. They translate to a very small
negative Poisson ratio between the soft $\vec{a_1}$ in-layer
direction and the $\vec{a_3}$ direction normal to the layers. The
Poisson ratio between the hard $\vec{a_2}$ in-layer direction and
the $\vec{a_3}$ direction is also very small in magnitude, but
changes sign near $a_2=0$. This definition of the Poisson ratio in
the bulk differs from the ``Poisson ratio'' in a phosphorene
monolayer, which relates the monolayer thickness to the in-layer
strain and finds a negative value for that
quantity~\cite{jiang2014nc}.

With the optimum value of the inter-layer spacing
$a_{3,opt}(a_1,a_2)$ for the different strain combinations at
hand, we have calculated the strain energy ${\Delta}E$ as a
function of ${\epsilon}(a_1)$ and ${\epsilon}(a_2)$ and present
the results in Fig.~\ref{fig1}(c). The prominently elliptical
shape of the isoenergetic contours is another manifestation of the
elastic anisotropy in the system. The observed tilt of the
elliptical axes from the horizontal and vertical direction
indicates a positive Poisson ratio $\nu_{21}=-d\epsilon{\rm
(a_2)}/d\epsilon{\rm (a_1)}=0.19$ within the phosphorene plane,
indicating that stretching in one (in-plane) direction results in
a lattice contraction in the normal (in-plane) direction. We find
the lattice to be rather soft with respect to in-plane
compression, since stretching by 2\% even along the stiffer $a_2$
direction requires an energy investment of only ${\approx}0.06$~eV
per unit cell.

\begin{figure}[t]
\includegraphics[width=0.7\columnwidth]{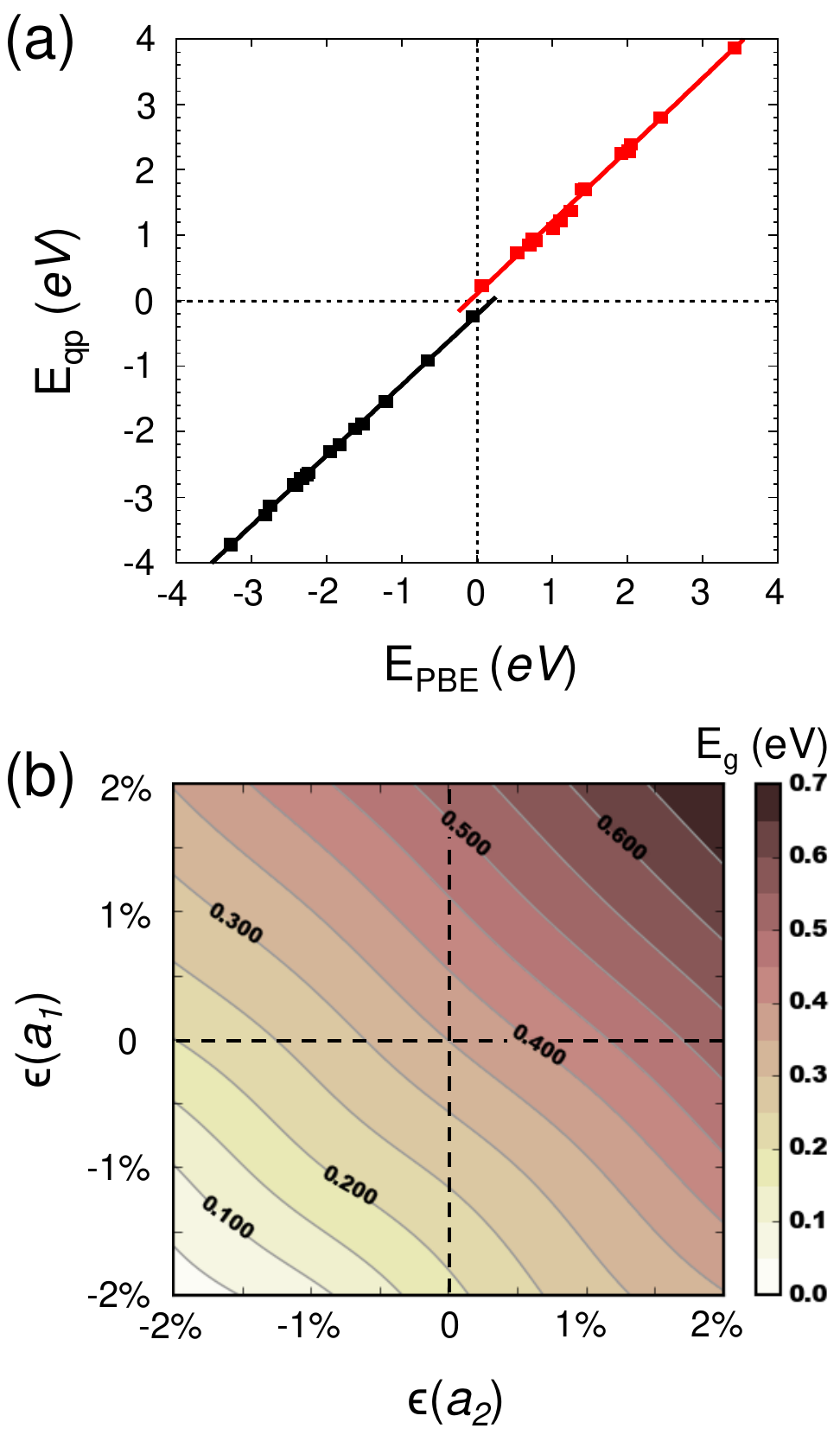}
\caption{(a) Correlation between quasiparticle (GW) energies
$E_{\rm qp}$ and Kohn-Sham energy values $E_{\rm PBE}$, obtained
using the DFT-PBE functional, for states along high symmetry lines
in the Brillouin zone. The straight lines in the valence and
conduction band regions are drawn to guide the eye. The results
represent bulk black phosphorus subject to strain values
$\epsilon(a_1)=1\%$ and $\epsilon(a_2)=0$. The Fermi level is at
zero energy. (b) Dependence of the quasiparticle (GW) electronic
band gap $E_g$ on the in-layer strain applied along the
$a_1$ and $a_2$ directions.%
\label{fig3} }
\end{figure}

Results of our calculations for the electronic band structure and
density of states (DOS) of unstrained and strained bulk black
phosphorus are shown in Fig.~\ref{fig2}. The DFT-PBE results,
represented by the dashed red lines in Fig.~\ref{fig2}(a), predict
an extremely small direct fundamental band gap value $E_g(\rm
PBE){\approx}0.05$~eV for unstrained bulk black phosphorus. The
band gap is becoming larger in the strained structures by
stretching along the $a_1$ direction. As mentioned before, the DFT
results are known to substantially underestimate the band gap in
semiconductors. Band structure results obtained using the more
proper GW approach are represented by the solid black lines in
Fig.~\ref{fig2}(a). These data suggest a larger quasiparticle band
gap $E_g(\rm qp){\approx}0.35$~eV in unstrained bulk black
phosphorus, very close to published value based on GW
calculations~{\cite{{Tran2014},{SWS:GW-TBA:PhysRevB.92.085419}}}
and to the observed value~\cite{bulkbp} of ${\approx}0.33$~eV.

Differences between quasi-particle spectra $E_{\rm qp}$ and DFT
band structure results $E_{\rm PBE}$ are summarized in
Fig.~\ref{fig3}(a). As shown previously~\cite{{GW86},{Zhang1989}},
the self-energy (or GW) correction is roughly represented by a
``scissor operator'', which would shift DFT-based valence states
rigidly down and conduction states rigidly up by
${\lesssim}0.2$~eV in bulk black phosphorus, thus opening up the
fundamental band gap.

A more precise comparison between the quasiparticle spectra and
DFT eigenvalues reveals that the difference $E_{\rm qp}-E_{\rm
PBE}$ does depend on the energy, but is independent of the crystal
momentum {\bf k}. We considered black phosphorus stretched by 1\%
along the soft $a_1$ direction, which has a nonzero band gap in
DFT-PBE, and displayed the correlation between quasiparticle
energies $E_{\rm qp}$ and corresponding DFT eigenvalues $E_{\rm
PBE}$ at selected high symmetry points in the Brillouin zone in
Fig.~\ref{fig3}(a). Besides the discontinuity at the Fermi level,
we found that the quasiparticle energies display a linear
relationship with DFT eigenvalues, given by
\begin{eqnarray}
E_{{\rm qp}}{\rm (CB)}&=&1.10{\times}
        E_{{\rm PBE}}{\rm (CB)}+0.11~{\rm eV}
       \label{eqn:gwpbe1}\\
E_{{\rm qp}}{\rm (VB)}&=&1.07{\times}
       E_{{\rm PBE}}{\rm (VB)}-0.18~{\rm eV}\;.
       \label{eqn:gwpbe2}%
\end{eqnarray}
Assuming that the Fermi level defines zero energy, the linear
relationship between $E_{\rm qp}$ and $E_{\rm PBE}$ is slightly
different in the conduction band (CB) region, identified by
$E_{\rm qp}>0$, and the valence band region, identified by $E_{\rm
qp}<0$. Since GW energies have been calculated only at a few
k-points, we have used expressions (\ref{eqn:gwpbe1}) and
(\ref{eqn:gwpbe2}) to generate the continuous GW band structure
shown in Fig.~\ref{fig2}.

Comparing DFT and GW values at different strains, we found that
the modulation of the band gap ${\Delta}E_g(\rm
PBE){\approx}{\Delta}E_g(\rm qp)$ is the same up to
${\lesssim}0.01$~eV in the strain range studied here. With the
quasiparticle band gap of unstrained black phosphorus at hand, we
thus can deduce the quasiparticle band gap in phosphorus subject
to different strain values ${\epsilon}(a_1)$ and ${\epsilon}(a_2)$
combining the DFT band gap values $E_g$(PBE) with
Eqs.~(\ref{eqn:gwpbe1}) and (\ref{eqn:gwpbe2}). Our results, based
on a cubic spline interpolation of a $5{\times}5$ grid of data
points, are shown in Fig.~\ref{fig3}(b).

Our results indicate that, within the range
$|\epsilon|{\lesssim}2\%$ of strain applied along the $a_1$ and
$a_2$ direction, the band gap $E_g$ of bulk black phosphorus
varies smoothly between $0.05$~eV and $0.70$~eV. Independent of
the strain direction, $E_g$ increases upon stretching and
decreases upon compression. Noticing the nearly equidistant
spacing between the contour lines in Fig.~\ref{fig3}(b), we can
extrapolate to larger strains and expect the band gap to close at
compressive strains along both $a_1$ and $a_2$ directions
exceeding 2\%.

So far, much less attention has been paid to the moderate 0.3~eV
band gap of bulk BP than to 2~eV wide direct band gap of a
phosphorene monolayer~\cite{DT229}. The inter-layer coupling in
bulk BP, which is responsible for the large difference in the band
gap, displays a nontrivial character~\cite{DT250}. Therefore, it
has not been clear {\em a priori} if previously reached
conclusions for the band gap dependence on uniaxial in-layer
strain in a phosphorene monolayer will also be applicable for the
bulk system. There is an independent, more practical concern about
band gap modulation in BP systems by strain. Pristine phosphorene
monolayers and few-layer systems are very unstable under ambient
conditions~\cite{{water_bpox2015},{Hersam14},{DT249}} and must be
capped, typically by sandwiching in-between inert h-BN layers to
remain useful~\cite{ChenWuWuEtAl2015}. It is unclear if applying
in-layer strain will not destroy the rigid capping layer before
reaching desirable strain values in the enclosed phosphorene. This
limitation applies to a much lesser degree to bulk black
phosphorus, which is chemically much more stable and thus can be
handled more easily in the experiment.

The sensitivity of the band gap to the in-layer strain suggests to
use bulk or multi-layer black phosphorus in infrared devices
tunable by strain. Optical measurements should be able to reveal
the band gap value discussed here, since observed optical spectra
should not be modified by excitonic states due to the negligibly
small exciton binding energy in bulk black
phosphorus~\cite{Tran2014}.


In summary, we have studied theoretically the structural and
electronic response of layered bulk black phosphorus to in-layer
strain. {\em Ab initio} density functional theory (DFT)
calculations reveal that the strain energy and interlayer spacing
display a strong anisotropy with respect to the uniaxial strain
direction. To correctly describe the dependence of the fundamental
band gap on strain, we used the computationally more involved GW
quasiparticle approach that is free of parameters and superior to
DFT studies, which are known to underestimate gap energies. We
found that the main difference between GW quasiparticle energies
$E_{\rm qp}$  and DFT eigenvalues $E_{\rm PBE}$ is a discontinuity
at the Fermi level and have identified the relationship between
$E_{\rm qp}$ and $E_{\rm PBE}$ in the valence and conduction band
regions. Similar to a phosphorene monolayer, we found that the
band gap depends sensitively on the in-layer strain and even
vanishes at compressive strain values exceeding ${\approx}2$\%,
thus suggesting a possible application of black P in
strain-controlled infrared devices.

\begin{acknowledgments}
J.G. and D.T. acknowledge the hospitality of Fudan University,
where this study was initiated, and useful discussion with Yuanbo
Zhang. We are also grateful to Garrett B. King for adapting the
contour plot code and for useful discussions. J.G. and D.T.
acknowledge financial support by the NSF/AFOSR EFRI 2-DARE grant
number EFMA-1433459. W.S. and L.Y. acknowledge support by the NSF
grant number EFRI-2DARE-1542815 and NSF CAREER grant number
DMR-1455346. Computational resources have been provided by the
Michigan State University High Performance Computing Center and
the Stampede of Teragrid at the Texas Advanced Computing Center
(TACC) through XSEDE.
\end{acknowledgments}

%

\end{document}